\documentclass[aps,pra,twocolumn,superscriptaddress]{revtex4-1}

\usepackage[utf8]{inputenc}
\usepackage{amsmath}
\usepackage{cases}
\usepackage{graphicx}
\usepackage{mathtools} % celtic 
\usepackage[T1]{fontenc} % celtic
\usepackage[utf8]{inputenc} % celtic
\usepackage{amsfonts} %celtichttps://www.overleaf.com/1382372365vcxzgscjzptw
\usepackage{bm}
\usepackage{dsfont}

\usepackage{enumitem}
\usepackage{multirow}
\usepackage{tabularx}
\usepackage{longtable}
\usepackage{float}
\usepackage{array}
\newcolumntype{H}{>{\setbox0=\hbox\bgroup}c<{\egroup}@{}}

\usepackage{acronym}
\usepackage{hyperref}
\usepackage[capitalise,nameinlink]{cleveref}
\usepackage[usenames,dvipsnames]{xcolor}
\hypersetup{
    colorlinks=true,       % false: boxed links; true: colored links
    linkcolor=Maroon,          % color of internal links (change box color with linkbordercolor)
    citecolor=OliveGreen,        % color of links to bibliography
    filecolor=magenta,      % color of file links
    urlcolor=Blue           % color of external links
}
\crefname{ansatz}{Ansatz.}{Ansatzes.}

\def\ket#1{\left|#1\right\rangle}

\newcommand{\PNNL}{%
    \affiliation{%
        William R. Wiley Environmental Molecular Sciences Laboratory, Battelle, 
        Pacific Northwest National Laboratory, K8-91, P.O. Box 999, Richland WA 99352, USA}}
\newcommand{\UFCH}{
\affiliation{%
J. Heyrovsk\'{y} Institute of Physical Chemistry,
Academy of Sciences of the Czech Republic,
Dolej\v{s}kova 3, CZ-18223 Praha, Czech Republic 
}}

\newcommand{\FJFI}{
\affiliation{%
Faculty of Nuclear Science and Physical Engineering, Technical University of Prague,
B\v{r}ehov\'{a} 7, CZ-11519 Praha, Czech Republic
}}

\begin{document}

\title{Variational Quantum Eigensolver for Approximate  Diagonalization of Downfolded Hamiltonians using  Generalized Unitary Coupled Cluster Ansatz}

 \author{Nicholas P. Bauman}
 \email{nicholas.bauman@pnnl.gov}
 \PNNL
 \author{Jaroslav Chl\'{a}dek}
 \email{chladjar@fjfi.cvut.cz}
 \FJFI
 \UFCH
 \author{Libor Veis}
 \email{libor.veis@jh-inst.cas.cz}
 \UFCH
 \author{Ji\v{r}\'{\i} Pittner}
 \email{jiri.pittner@jh-inst.cas.cz}
 \UFCH
 \author{Karol Kowalski}
 \email{karol.kowalski@pnnl.gov}
 \PNNL
\date{January  2020}

\begin{abstract}
In this paper we discuss the utilization  of Variational Quantum Solver (VQE) and recently introduced   Generalized Unitary Coupled Cluster (GUCC) formalism for the  diagonalization of  downfolded/effective Hamiltonians in active spaces. In addition to effective Hamiltonians defined by the downfolding of a subset of virtual orbitals we also consider their form defined by freezing core orbitals, which enables us to deal with larger systems. We also consider various solvers to identify solutions of the GUCC equations. We use N$_2$, H$_2$O, and  C$_2$H$_4$, and benchmark systems to illustrate the performance of the combined framework. 
\end{abstract}

\maketitle

\section{Introduction}
As originally envisaged by Feynman,\cite{feynman1982simulating} complex quantum system simulations should be performed using computational devices (quantum computers) operating based on principles of quantum mechanics and its stochastic character. The last few decades have been  marked by the intensive development of quantum algorithms designed to tackle many-body problems characterized by complexities  which cannot be effectively addressed even by employing computational power of near-exa-scale classical computer architectures. These difficulties originate in the necessity of   systematic inclusion of  collective many-body effects  that quickly leads to unsurmountable computational barriers that preclude in many cases predictive simulations of strongly correlated systems in chemistry and material sciences. Quantum computing brings the hope that these problems can be overcome and reliable simulations of quantum systems can be performed in finite times.   Another advantage of certain quantum computing algorithms (for example, Quantum Phase Estimation (QPE) \cite{luis1996optimum, cleve1998quantum,berry2007efficient,childs2010relationship,seeley2012bravyi,
wecker2015progress,haner2016high,poulin2017fast} 
algorithm) is the fact that they  significantly reduce (although it does not entirely eliminate) the bias of the conventional approximate approaches, typically associated with the choice of the broadly understood  zeroth-order approximations and  levels of accuracy,  and at least theoretically are capable of providing exact results at least in the stochastic sense.

The development of quantum algorithms for quantum chemistry  have permeated several simulations including ground-state calculations,\cite{cao2019quantum,mcardle2020quantum} excited-state modeling for low-lying valence excited states \cite{wang2008quantum,PhysRevX.8.011021,mcclean2017hybrid,
low2019q,higgott2019variational,veis2010} 
and core-level transitions,\cite{bauman2020towards} phase estimation for relativisitc and non-Born-Oppenheimer Hamiltonians, \cite{veis2012,veis2016_nbo} imaginary time evolution and  quantum Lanczos algorithms for thermal states, \cite{mcardle2019variational,motta2020determining,stair2020multireference} quantum filter diagonalization, \cite{parrish2019quantum} quantum inverse iterations, \cite{kyriienko2020quantum}, quantum power methods \cite{seki2020quantum}, and efficient preparation of correlated fermionic states \cite{veis2014,veis2019}.
Nowadays, two major classes of quantum algorithms are being  intensively developed with the aim of applying them to realistic systems on current noisy intermediate-scale quantum (NISQ) devices: (1) mentioned earlier family of  QPE algorithms, and (2) broad class of Variational Quantum Eigensolver (VQE) methods.\cite{peruzzo2014variational,mcclean2016theory,romero2018strategies,PhysRevA.95.020501,Kandala2017,kandala2018extending,PhysRevX.8.011021,huggins2020non,guzik2019review}
These two algorithms are driven  by different computational strategies and characterized by different resource requirements. While VQE draws heavily on the explicit form of the wave function Ansatz $|\Psi(\vec{\theta})\rangle$ with the polynomial number of variational parameters $\theta_i$, the QPE 
utilizes unitary evolution operator $e^{-itH}|\Phi\rangle$ for judiciously chosen initial guess $|\Phi\rangle$ for the exact wave function. 

The applicability of VQE and QPE algorithms  is either limited by  the number of variational variables and the numbers of necessary measurements  or by the circuit depths, respectively, which was extensively discussed in the literature (see for example Ref. \cite{von2020quantum}).   For the VQE formalism, the unitary coupled cluster (UCC)\cite{pal1984use,hoffmann1988unitary,kutzelnigg1991error,
unitary1,unitary2,sur2008relativistic,cooper2010benchmark,
harsha2018difference244,evangelista2019exact} 
expansion (as well as it generalized UCC variant (GUCC)
\cite{lee2018generalized})
is most frequently used form for the  wave function Ansatz, where the impact of higher-rank excitations
and the role of the orbital choice \cite{sokolov2020quantum,PhysRevResearch.2.033421} have been explored.
A non-orthogonal multi-reference variant of VQE logical Ansatz  (NOVQE) \cite{huggins2020non} has also been recently tested and verified for molecules characterized by the presence of strong correlation effects. 
On the other hand, the progress achieved in the development of QPE formalism should be mostly attributed to the development of novel Trotterization  and Taylor expansion techniques \cite{trotter1959product,suzuki1990fractal,wiebe2010higher,poulin2014trotter,berry2015simulating} (or approximations the exponential form of the unitary operator in general) and new representations of the second-quantized form of the many-body Hamiltonians  stemming from the double decomposition of tensors describing pairwise interactions.\cite{peng2017highly,motta2018low,von2020quantum} 
Using these techniques it was demonstrated that the overall number of terms entering the many-body Hamiltonian can be significantly reduced, which leads to a significant reduction in the gate depth of the corresponding quantum circuit. 

In the last few years several methodologies have been explored to reduced the dimensionality of many-body Hamiltonians used in quantum computing. This includes theory of transcorrelated Hamiltonians, canonical transformation theory, and subspace expansions.\cite{,motta2020quantum,mcardle2020improving,takeshita2020increasing} 
In the case of the downfolding methods \cite{bauman2019downfolding,downfolding2020t} one can construct active-space representation of  effective/downfolded Hamiltonians 
which integrate out high-energy Fermionic degrees of freedom while being capable of reproducing exact energy of quantum systems. These formulations based on the double unitary CC (DUCC) Ansatz have recently been used in the context of quantum computing for ground- and excited-state problems showing a promise in the recovering the exact energy of the full problem 
using limited quantum resources. 

For further development of various downfolding  methods a crucial role is played by the possibility of integrating them with various Ansatzen that can provide optimal utilization of quantum VQE algorithm. For this reason, in this paper we propse a variant of the  VQE algorithm that utilize generalized UCC Ansatz   to diagonalize downfolded Hamiltonian in active spaces. 
We will refer the resulting formalism to as the downfolding VQE algorithm with GUCC form of trial wave function or DVQE-GUCC for short. 
We also evaluate the performance of various classical VQE optimizers for systems characterized by various complexity levels of ground-state electronic wave functions. In particular, we compare DVQE-GUCC results for three benchmark systems (H$_2$O, C$_2$H$_4$, and N$_2$) with the exact results obtained by the full configuration interaction (FCI) method.

\section{CC downfolding techniques}
In Ref.~\cite{bauman2019downfolding}, we introduced the unitary extension of the sub-system embedding sub-algebra CC approach (SES-CC)~\cite{safkk} which utilizes the double unitary CC expansion. The Hermitian form of CC downfolding is predicated on a special form of wave function expansion given by double unitary coupled cluster Ansatz (DUCC) given by the expressions (for more details see Ref.\cite{bauman2019downfolding}) 
\begin{equation}
	|\Psi\rangle=e^{\sigma_{\rm ext}} e^{\sigma_{\rm int}}|\Phi\rangle;,
\label{ducc1}
\end{equation}
where $\sigma_{\rm int}$ and $\sigma_{\rm ext}$ are the general type anti-Hermitian operators 
\begin{eqnarray}
\sigma_{\rm int}^{\dagger}&=&-\sigma_{\rm int} \;,
\label{gahint} \\
\sigma_{\rm ext}^{\dagger}&=&-\sigma_{\rm ext}
\label{gahext}
\end{eqnarray}
defined by 
%excitations/de-excitations 
amplitudes defining action 
within and outside of the pre-defined active space, respectively, i.e.,  the amplitudes defining the $\sigma_{\rm ext}$ operator must carry at least one inactive spin-orbital index whereas all amplitudes defining the $\sigma_{\rm int}$ operator carry active spinorbital indices only. In Eq.(\ref{ducc1}), $|\Phi\rangle $ designates properly chosen reference function (usually chosen as a Hartree-Fock (HF) Slater determinant). The exactness of the expansion \ref{ducc1}
has been recently discussed in Ref.\cite{downfolding2020t} where it was also shown that the standard UCC expansions can provide a basic approximation of the exact $\sigma_{\rm int}$ and $\sigma_{\rm ext}$
operators, i.e.,
\begin{eqnarray}
\sigma_{\rm int} \simeq T_{\rm int} - T_{\rm int}^{\dagger}
\;, \label{uccint} \\
\sigma_{\rm ext} \simeq T_{\rm ext} - T_{\rm ext}^{\dagger}
\;, \label{uccext}
\end{eqnarray}
where $T_{\rm int}$ and $T_{\rm ext}$ are single-reference-type internal and external cluster amplitudes (in the sense defined above).

Introducing \ref{ducc1} into Schr\"odinger's equation,  premultiplying both sides by $e^{-\sigma_{\rm int}}e^{-\sigma_{\rm ext}}$, and projecting the resulting equations onto subspace of excited determinants, with respect to the reference function $|\Phi\rangle$, defined by a projection operator $Q$ and onto reference function (with corresponding projection operator $P=|\Phi\rangle\langle\Phi|$,
one obtains equations for cluster amplitudes and the 
corresponding energy, i.e., 
\begin{eqnarray}
        Qe^{-\sigma_{\rm int}}e^{-\sigma_{\rm ext}} H e^{\sigma_{\rm ext}}e^{\sigma_{\rm int}}|\Phi\rangle &=& 0,
\label{uccd2eq} \\
        \langle\Phi|e^{-\sigma_{\rm int}}e^{-\sigma_{\rm ext}} H e^{\sigma_{\rm ext}}e^{\sigma_{\rm int}} |\Phi\rangle &=& E,
\label{uccd2ene}
\end{eqnarray}
Using the {\it Campbell-Baker-Hausdorff} formula, the above formulas can be put in explicitly connected form 
\begin{eqnarray}
        Q ( H e^{\sigma_{\rm ext}}e^{\sigma_{\rm int}})_C|\Phi\rangle &=& 0,
\label{uccd2eqx} \\
        \langle\Phi|( H e^{\sigma_{\rm ext}}e^{\sigma_{\rm int}})_C |\Phi\rangle &=& E,
\label{uccd2enex}
\end{eqnarray}
where subscript "C" stands for a connected part of a given operator expression. In contrast to  standard CC formulations, expansions in above equations are non-terminating. 

For the sake of simplicity, in the following analysis  we assume  the case of the exact limit ($\sigma_{\rm int}$ and 
$\sigma_{\rm ext}$ include all possible excitations). 
In Ref.~\cite{bauman2019downfolding}, we showed that when $\sigma_{\rm int}$ contains all possible excitations/de-excitations within the 
active space, the energy of the system~\cref{uccd2ene} can be obtained by diagonalizing
the DUCC effective Hamiltonian
\begin{equation}
        \overline{H}_{\rm eff}^{\rm (DUCC)} e^{\sigma_{\rm int}} \ket{\Phi} = E e^{\sigma_{\rm int}}\ket{\Phi},
\label{duccstep2}
\end{equation}
where $\overline{H}_{\rm eff}^{\rm (DUCC)}$ is defined as
\begin{equation}
        \overline{H}_{\rm eff}^{\rm (DUCC)} = (P+Q_{\rm int}) \overline{H}_{\rm ext}^{\rm DUCC} (P+Q_{\rm int})
\label{equivducc}
\end{equation}
and 
\begin{equation}
        \overline{H}_{\rm ext}^{\rm DUCC} =e^{-\sigma_{\rm ext}}H e^{\sigma_{\rm ext}}.
\label{duccexth}
\end{equation}
In the above equations, the   $e^{\sigma_{\rm int}}|\Phi\rangle$ vector defines the corresponding eigenvector. To prove this property, it is sufficient to introduce the resolution of identity $e^{\sigma_{\rm 
int}}e^{-\sigma_{\rm int}}$ to the left of the $\overline{H}_{\rm ext}^{\rm DUCC} $ operator in 
\begin{equation}
	(P+Q_{\rm int}) \overline{H}_{\rm ext}^{\rm DUCC} e^{\sigma_{\rm int}} |\Phi\rangle = E
    (P+Q_{\rm int}) e^{\sigma_{\rm int}}|\Phi\rangle\;,
\label{duccstep1}
\end{equation}
where we employed the fact that 
\begin{equation}
(P+Q_{\rm int}) e^{\sigma_{\rm int}}|\Phi\rangle =
e^{\sigma_{\rm int}}|\Phi\rangle \;,
\label{pqipro}
\end{equation}
and to notice that $e^{-\sigma_{\rm int}}\overline{H}_{\rm ext}^{\rm 
DUCC} e^{\sigma_{\rm int}}=e^{-\sigma_{\rm int}}e^{-\sigma_{\rm ext}} H e^{\sigma_{\rm 
ext}}e^{\sigma_{\rm int}}$. 
%Next, in analogy to Eqs. (\ref{analysis1}) and 
%(\ref{llineq}), Eq. (\ref{equivducc}) can be represented as 
Next, using matrix representation of the $\sigma_{\rm int}$ operator in the CAS space, denoted 
as $\bm{\sigma}_{\rm int}$, this equation can be re-written as 
\begin{equation}
	[e^{\bm{\sigma}_{\rm int}}] [\bm{y}] = 0 \;,
\label{llineq2}
\end{equation}
where the first component of the $[\bm{y}]$ vector is equivalent to 
$\langle\Phi|e^{-\sigma_{\rm int}}e^{-\sigma_{\rm ext}} H 
e^{\sigma_{\rm ext}}e^{\sigma_{\rm int}} |\Phi\rangle-E$
while the remaining components correspond to projections of 
$e^{-\sigma_{\rm int}}e^{-\sigma_{\rm ext}} H 
e^{\sigma_{\rm ext}}e^{\sigma_{\rm int}} |\Phi\rangle$
onto excited configurations belonging to $Q_{\rm int}$. 
The $[e^{\bm{\sigma}_{\rm int}}]$ matrix is also  non-singular, which is a consequence of the formula 
\begin{equation}
	{\rm det}(e^{\bm{\sigma}_{\rm int}})=e^{{\rm Tr}(\bm{\sigma}_{\rm int})} =1
\label{det1}
\end{equation}
and the anti-Hermitian character of the $\bm{\sigma}_{\rm int}$
matrix, i.e., ${\rm Tr}(\bm{\sigma}_{\rm int})=0$
(where real character of $\sigma_{\rm int}$ cluster amplitudes is assumed).
Given the non-singular character 
of the $[e^{\bm{\sigma}_{\rm int}}]$ matrix 
(see also  Ref.~\cite{bauman2019downfolding}), this proves the equivalence of these two representations. 

The discussed variant of double unitary CC expansion extends properties of single reference CC sub-system embedding sub-algebra (SES-CC) \cite{safkk} to the unitary CC case. In analogy to SES-CC case, the  DUCC expansion based on Ansatz (\ref{ducc1})  leads to the rigorous decoupling of fermionic degrees of freedom in the effective Hamiltonian. As discussed in earlier papers these degrees of freedom can be associated with energy,\cite{bauman2019downfolding} time,\cite{kowalski2020sub} and spatial scales.\cite{bauman2019downfolding}

In the following part of the paper we will study the approximate form of the $\overline{H}_{\rm eff}^{\rm DUCC}$ operator. 
In analogy to Ref.\cite{bauman2019downfolding} we will use the approximate form of the $\overline{H}_{\rm eff}^{\rm DUCC}$, which is consistent with the second order  energy expansion, i.e.,
\begin{widetext}
\begin{equation}
\label{DUCChamiltonian}
 \overline{H}_{\rm eff}^{\rm DUCC}\simeq (P+Q_{\rm int}) (H+[H_N,\sigma_{\rm ext}] +\frac{1}{2!}[[F_N,\sigma_{\rm ext}],\sigma_{\rm ext}])(P+Q_{\rm int}) \;,
\end{equation}
\end{widetext}
where the  cluster operator $\sigma_{\rm ext}$ 
is approximated as 
\begin{equation}
    \sigma_{\rm ext} \simeq T_{\rm ext,1} + T_{\rm ext,2} - T_{\rm ext,1}^{\dagger} - T_{\rm ext,2}^{\dagger} \;,
\label{approx1}
\end{equation}
where $T_{\rm ext,1}$ and $T_{\rm ext,2}$ are external pars of the CCSD $T_1$ and $T_2$ operators. 
Moreover, 
$\overline{H}_{\rm eff}^{\rm DUCC}$ is approximated by one- and two-body effects discussed in Ref.\cite{bauman2019downfolding}.

\section{VQE algorithm and generalized unitary CCSD Ansatz}
Although the quantum phase estimation algorithm \cite{luis1996optimum, cleve1998quantum,berry2007efficient,childs2010relationship,seeley2012bravyi,
wecker2015progress,haner2016high,poulin2017fast} 
%and its iterative variants \cite{} 
provide efficient means to obtain
the exact full CI solution on a quantum computer, as far as presently understood, it requires
large number of qubits, with quantum error correction, and long coherence times, which are (except for tiny model systems) beyond the reach of present quantum hardware. A promising alternative for near-term quantum devices is the variational quantum eigensolver (VQE) approach \cite{peruzzo2014variational,mcclean2016theory,romero2018strategies,PhysRevA.95.020501,Kandala2017,kandala2018extending,PhysRevX.8.011021,huggins2020non,guzik2019review}, which combines classical
variational energy minimization over normalized trial wave functions $\Psi(\vartheta)$ parameterized by a set of variables $\vartheta$
\begin{equation}
\label{varmin}
    E = \min_{\vartheta} \, \langle \Psi(\vartheta)|H|\Psi(\vartheta)\rangle
\end{equation}
with computation of the Hamiltonian matrix elements on a quantum device. 
The ``bare'' second-quantized molecular Hamiltonian can be expressed as a sum of one- and two-body terms, formally
\begin{equation}
\label{Hterms}
    H = \sum_X h_X \hat{X}
\end{equation}
where $\hat{X}$ is a string of up to two creation and two annihilation operators.
The DUCC effective Hamiltonian (\ref{DUCChamiltonian}) can be approximated in the same form, when higher-body terms are neglected.
In order to minimize the complexity of the quantum circuit, expectation values of each
term in the summation (\ref{Hterms}) are evaluated separately and results added classically.
The advantage of this approach consists in the fact that evaluation of these matrix elements for many trial wave functions, which would lead to exponential complexity on a classical computer, can be efficiently performed on a quantum computer
with a moderate circuit complexity and coherence time requirements \cite{whaley2019vqe}.

A popular wave function ansatz is the exponential CC form
\begin{equation}
\label{expansatz}
    \Psi(t) = e^T |\Phi\rangle
\end{equation}
where at the singles and doubles truncation level the cluster operator has the form
\begin{equation}
    T= T_1 + T_2
\end{equation}
with
\begin{eqnarray}
T_1 &=& \sum_{ia} t_i^a a_a^\dagger a_i\\
T_2 &=& \sum_{i<j, a<b} t_{ij}^{ab} a_a^\dagger a_b^\dagger a_j a_i
\end{eqnarray}
where we adopt the convention of $i,j,\ldots$ running over occupied spinorbitals and $a,b,\ldots$ running over the virtual ones.
In the traditional CC method \cite{shavitt-bartlett} this ansatz is not employed in the variational minimization (\ref{varmin}), since that yields infinite diagrammatic expansions, which are inconvenient to evaluate and efficiently approximate on a classical computer.
A projection technique is employed instead, which leads to closed explicit algebraic equations
for the parameters $t_i^a$, $t_{ij}^{ab}$. For the same reason, the unitary coupled cluster ansatz (UCCSD)
\begin{equation}
\label{uccansatz}
    \Psi(t) = e^{T-T^\dagger} |\Phi\rangle
\end{equation}
was never widely employed in mainstream quantum chemistry, although considerable effort has been dedicated to development of computationally efficient approximations \cite{Taube2006}.
On the other hand, on a quantum computer the unitary ansatz is clearly advantageous as the wave function parametrization then straightforwardly and efficiently translates to a (unitary) quantum circuit.
However, it still bears the main disadvantage of the single-reference CC method, that it is based on truncated excitations from a single reference determinant $\Phi$, which hampers it application to strongly correlated (multiconfigurational) systems.

A possible generalization of the CCSD ansatz (\ref{expansatz}) consists in including both excitations and deexcitations into the cluster operator, which thus attains the CCGSD form
\begin{eqnarray}
\label{ccgsd}
T_1 &=& \sum_{pq} t^p_q a_p^\dagger a_q \\
T_2 &=& \sum_{p<q, r<s} t^{pq}_{rs} a_p^\dagger a_q^\dagger a_s a_r
\end{eqnarray}
where the indices $p,q,r,s$ run over all spinorbitals and the reference to a Fermi vaccum
is thus eliminated from the cluster operator. Moreover, the ``reference'' wave function $\Phi$
in (\ref{expansatz}) need not be a single Slater determinant, but any state that can be efficiently prepared on a quantum computer.
In the early 2000s this ansatz was subject of an active debate in the literature \cite{nooijen-general,head-gordon-general,nakatsuji1,nakatsuji2,nakatsuji3,nakatsuji-davidson,piecuch_2003,davidson2003,ronen2003,mazziotti2004,nakatsuji2004,kutzelnigg2004gccsd,kutzelnigg2005gccsd}, since
Nooijen conjectured \cite{nooijen-general} that the ground state  solution of a two-body Hamiltonian can be exactly expresed in this form. This conjecture was later disproved by Kutzelnigg and Mukherjee based on Lie-algebraic arguments \cite{kutzelnigg2004gccsd,kutzelnigg2005gccsd}
(after excluding solutions in the infinite $t$ limit where $e^T$ is a projection operator and precisely defining what is allowed to be the reference wave function $\Phi$).
Nevertheless, several studies in the meantime have shown, that even if not exact,
it yielded excellent numerical agreement compared to FCI \cite{head-gordon-general,piecuch_2003}. This has recently motivated its use in the VQE context \cite{whaley2019vqe}, in the unitary modification (UCCCGSD), when the CCGSD cluster operator (\ref{ccgsd}) is combined with the unitary ansatz (\ref{uccansatz}). Notice that since in the CCGSD
both excitations and deexcitations are included, the antihermitian linear combination $T-T^\dagger$ is also expressible in the CCGSD form (\ref{ccgsd}) where the unitarity manifests itself only in restrictions on the values of the amplitudes of mutually conjugated excitation operators, so that UCCGSD trial wave function has the same form as CCGSD, just with less free parameters, while the unitarity represents an advantage for the implementation in a quantum circuit. 
The mutual non-commutativity of the terms in the CCGSD cluster operator (\ref{ccgsd}) has two implications. Firstly, in the qunatum implementation it has to be approximated by a Trotter expansion. However, previous numerical study has achieved excellent accuracy even with very low number of Trotter steps \cite{whaley2019vqe}.
Secondly, it complicates the evaluation of the analytic gradient of the energy expectation value (\ref{varmin}), since one cannot straightforwardly apply the chain rule. This problem has
already been discussed and solved by Van Voorhis and Head-Gordon \cite{head-gordon-general} using the Wilcox identity \cite{wilcox}.

In our numerical study, we employed the DUCC effective Hamiltonian truncated to one- and two-body terms (\ref{DUCChamiltonian}) and performed the variational minimization (\ref{varmin}) using the BFGS method with analytic gradient. The CCGSD ansatz (\ref{ccgsd}) 
has been implemented using an in-house code based on full-CI expansion of the excitation operators and numerical evaluation of the exponential (\ref{uccansatz}) in a ``direct-CI'' style. We have employed several choices of the reference state, either a single Slater determinant or matrix produxt state (MPS) wave function from a density matrix renormalization group (DMRG) calculations with several small values of bond dimension. Also the effect of different initial guess of amplitudes
(zero, MP2, MPS) on the convergence rate has been investigated.

\section{Results and Discussion}
\begin{figure}[t]
	\includegraphics[trim= 1.15in 4.85in 4.75in 0.85in, clip,  width = 0.46\textwidth]{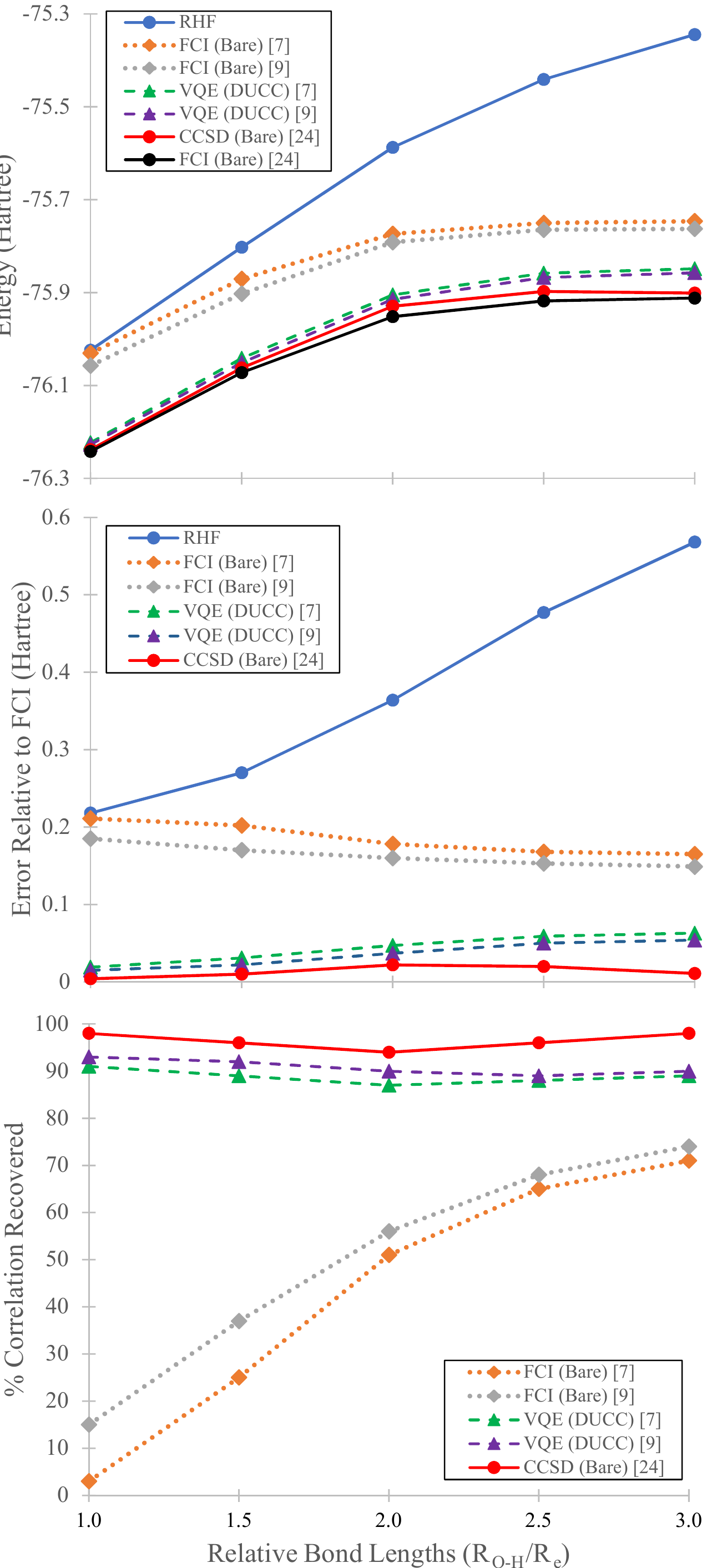}
	\caption{Total energies (top panel), errors (middle panel), and percentage of correlation recovered (bottom panel) for the $C_{2v}$-symmetric double dissociation of H$_{2}$O. The VQE and FCI results with the DUCC Hamiltonians agree to at least four digits as reported in Table \ref{H2O-Table}, so only one set of results are illustrated.}
	\label{fig:h2o}
\end{figure}

The $C_{2v}$-symmetric double dissociation of H$_{2}$O is a popular benchmark system to test
the accuracy methods under different regimes of static/dynamical correlation, since the degree 
of static correlation is systematically increased while stretching both bonds. At equilibrium,
simple truncation of the bare orbital space down to 7 or 9 orbitals disposes of over 85\% of the 
correlation energy and results in errors of 0.185--0.211 Hartree, relative to the full 
configuration interaction (FCI) values (see Table \ref{H2O-Table} and Figure \ref{fig:h2o}). 
As the bonds are stretched, the static correlation 
grows and the truncated orbital space recovers a larger percentage of the correlation. However, 
even when both bonds are stretched to three times the equilibrium length, over 25\% of the 
correlation is still missing, amounting to 0.149--0.165 Hartree errors. When correlation from outside 
of the active space is incorporated using the DUCC downfolding procedure, 87--93\% of the 
correlation is consistently captured along the double-dissociation pathway. Errors for the 7 
orbital active space are reduced to 0.019--0.063 Hartree, and are further reduced to 0.015--0.055
Hartree for the 9 orbital active space.

\begin{figure}[t]
	\includegraphics[trim= 1.85in 0.75in 1.75in 0.85in, clip,  width = 0.46\textwidth]{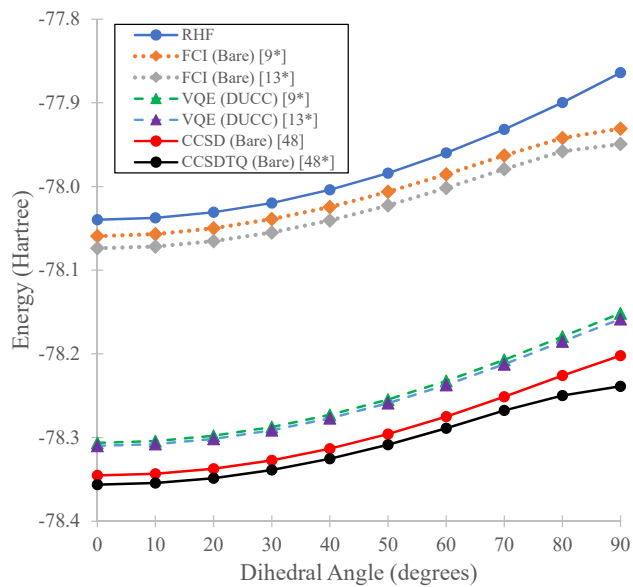}
	\caption{Total energies for the twisting of ethylene from the planar (0 degrees) to the twisted (90 degrees) confirmation.
	The VQE and FCI results with the DUCC Hamiltonians agree to at least four digits as reported in Table \ref{Ethylene-Table}, so only one set of results are illustrated. Numbers with an asterisk indicate that the atomic core orbitals were frozen in the corresponding calculation.}
	\label{fig:ethylene}
\end{figure}

We investigated the performance of the DUCC downfolding procedure
for the twisting of ethylene using two active spaces, 9 and 13 
orbitals. As seen in Table \ref{Ethylene-Table} and Figure \ref{fig:ethylene}, 
the bare Hamiltonian recovers 6--18\% of the correlation along the potential energy surface for the 9
orbital active space and 10--23\% for the 13 orbital active space, 
with errors of 0.28--0.30 and 0.29--0.31 Hartree, respectively, when 
compared with the benchmark CCSDTQ calculations. When the correlations from 
outside of the active space are included through the DUCC downfolding procedure, 76--86\%
of the correlation is recovered with either active space. In general, the DUCC Hamiltonians recover a 
more consistent portion of the correlation energy, with the largest errors occurring at the 90 degree
twisted confirmation. The CCSDTQ curve shows the flattening of the potential energy surface at 
this point. The CCSD calculations that underlies the downfolding procedure do not 
produce the correct curvature, which propagates into the corresponding DUCC Hamiltonians. This 
demonstrates the importance of the source and quality of the amplitudes defining the downfolding procedure
and the occasional need for amplitudes from higher-level methods. If the correlation effects inside and outside of the active space are well decoupled, the  
diagonalization of the DUCC Hamiltonian can overcome deficiencies 
in the underlying CCSD calculation, as we will see with N$_2$. It is important to mention that the 
active-space FCI calculations appear to have the correct curvature, but this is fortuitous as even CCSD
will have similar curvature in the same active space.

\begin{figure}[t]
	\includegraphics[trim= 1.60in 0.85in 1.6in 0.9in, clip,  width = 0.46\textwidth]{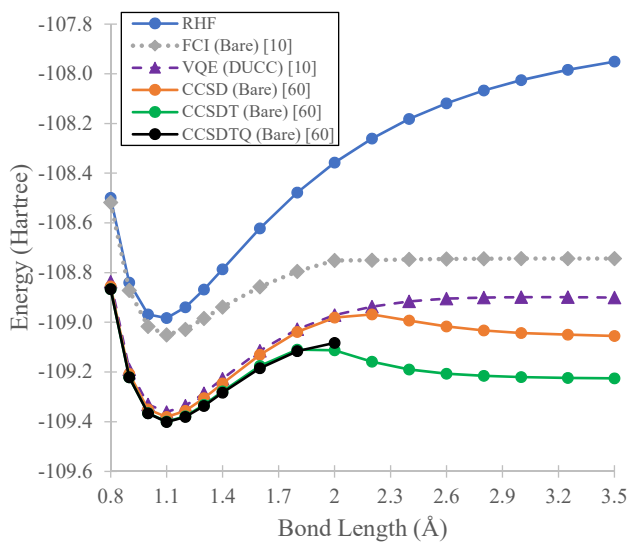}
	\caption{Total energies for the dissociation of N$_{2}$.
	The VQE and FCI results with the DUCC Hamiltonians agree to at least four digits as reported in Table \ref{N2-Table}, so only one set of results are illustrated.}
	\label{fig:N2}
\end{figure}

The bond breaking of N$_2$ is particularly difficult for conventional CC methods. Lower-order methods such as CCSD and 
even CCSDT have breakdowns that lead to nonphysical barriers (see Figure \ref{fig:N2}), and methods such as CCSDTQ will
run into difficulty converging at stretched bond lengths. The static and dynamical correlations are well decoupled
along the potential energy surface when defined by an active space of just 10 orbitals. However, the dynamical 
correlation missing from FCI calculations in the active space is significant, especially near equilibrium. When
those dynamical correlations are incorporated into the active space via the DUCC Hamiltonian, we recover
energies comparable to the full conventional CC methods, especially at equilibrium. Errors at stretched bond 
lengths will be higher, but considering that the underlying approach to the downfolding technique is CCSD, this is
rationalized. Once again, since the correlation effect in and out of the active space are well decoupled, the DUCC 
Hamiltonian results do not exhibit the breakdown that lower-order canonical CC methods, such as CCSD and CCSDT, have.
With amplitudes from a more suited method and longer expansions defining the DUCC Hamiltonian will further improve
upon the energies at the stretched bond lengths.

In tables \ref{table4} and \ref{table5} we present VQE results for N$_2$ molecule with bare Hamiltonian in STO-3G basis to investigate its convergence for different reference kets and inital guess of UGCCSD amplitudes.
We varied the reference ket state between HF and three different MPS states and used initial amplitudes either zero, or from MP2, or
from a CC analysis of the corresponding MPS wave function \cite{veis2016}.
As the convergence criterium for the BFGS quasi-Newton energy minimization we employed maximum amplitude change 1e-8 a.u. or
maximum gradient component 1e-5 a.u.

The first thing to note is that for both geometries, when HF is employed as a reference, the number of iterations 
before reaching convergence is relatively low compared to when the MPS8 or MPS16 are used. However, the final converged values 
have noticeable errors in energy compared to the FCI reference and relatively large errors in the  norm of difference from FCI and/or overlap with FCI. On the other hand, for MPS32 the number of iterations till convergence is lower and accuracy higher, which
can be explained by the fact that for such a small system the bond dimension 32 is already high enough for the reference being close
to the exact solution. Nevertheless, the DMRG energies corresponding to the MPS32 guess states have error still higher than 1 mHartree.  

Generally, MPS initial guess offers a better agreement with FCI, but this can comes at the cost of several more iterations before convergence in some cases. As the MPS gets larger, the accuracy generally increases toward FCI. However, we observed two ``pathological'' cases where the method converged to a local minimum
providing a solution with very high energy and very far from the ground state FCI wave function. This behavior seems
sporadic, and it might disappear for other values of convergence thresholds, but we would like to understand this further in future studies.
The choice of initial amplitudes played a little role in the convergence and achieved accuracy, except for the two aforementioned
cases when the minimization did not converge to the correct ground state with one value of initial amplitudes, while it did for all the other ones employed.

\section{Conclusions}

In this paper, we have presented the quantum approach for approximate diagonalization of DUCC down-folded Hamiltonians, which combines the variational quantum eigensolver with the generalized unitary coupled cluster ansatz and is aimed at near-term noisy quantum devices. 
We have demonstrated its capability to capture a large part of dynamical electron correlation missing in the active space (about 80 - 90 \% of total correlation) by numerical simulations of three benchmark systems of varying multireference character, namely N$_2$, H$_2$O, and C$_2$H$_4$. Last but not least, we have shown on the example of N$_2$, how the initial guesses in the form of matrix product states, which can be efficiently prepared on a quantum register, can improve the accuracy of the VQE-GUCC method for multireference systems.

\begin{table*}[ht]
\caption{Energies obtained with various methods using both bare and DUCC Hamiltonians for the double dissociation of H$_2$O$^\text{a}$}%{\footnotemark[1]}}
\begin{center}
\resizebox{6.5in}{!}{
\begin{tabular}{ c c c c H H c c c c c c c }
\hline\hline
             & HF & FCI & FCI & HF & HF & VQE & VQE & FCI & FCI & CCSD & FCI \\
Hamiltonian: & Bare & Bare & Bare & DUCC & DUCC & DUCC & DUCC & DUCC & DUCC & Bare & Bare \\
Orbitals: &  & [7] & [9] & [7] & [9] & [7] & [9] & [7] & [9] & [24] & [24] \\
\hline
Eq.     & 0.2178 & 0.2113 (3)  & 0.1846 (15) & 0.0253 (88) & 0.0429 (80) & 0.0189 (91) & 0.0148 (93) & 0.0189 (91) & 0.0148 (93) & 0.0037 (98) & -76.2419 \\
1.5*Eq. & 0.2700 & 0.2016 (25) & 0.1698 (37) & 0.0925 (66) & 0.1064 (61) & 0.0308 (89) & 0.0216 (92) & 0.0308 (89) & 0.0216 (92) & 0.0100 (96) & -76.0723 \\
2*Eq.   & 0.3640 & 0.1778 (51) & 0.1603 (56) & 0.2268 (38) & 0.2265 (38) & 0.0468 (87) & 0.0372 (90) & 0.0468 (87) & 0.0372 (90) & 0.0220 (94) & -75.9517 \\
2.5*Eq. & 0.4767 & 0.1679 (65) & 0.1532 (68) & 0.3870 (19) & 0.3671 (23) & 0.0595 (88) & 0.0503 (89) & 0.0595 (88) & 0.0503 (89) & 0.0203 (96) & -75.9180 \\
3*Eq.   & 0.5676 & 0.1654 (71) & 0.1491 (74) & 0.5106 (10) & 0.4781 (16) & 0.0633 (89) & 0.0545 (90) & 0.0633 (89) & 0.0545 (90) & 0.0108 (98) & -75.9119 \\
\hline\hline
\end{tabular}
}
\vspace{-1em}
\footnotetext[1]{
\setlength{\baselineskip}{1em} 
FCI energy values in the last column are in Hartree. All other quantities are errors relative to these energies. Values in parentheses are the percentages of correlation recovered.
}
\end{center}
\label{H2O-Table}
\end{table*}

\begin{table*}[ht]
\caption{Energies obtained with various methods using both bare and DUCC Hamiltonians 
for the twisting of ethylene.}
\begin{center}
% \resizebox{6.5in}{!}{%
\begin{tabular}{ c c H c H H c c c c H c H c c}
\hline\hline
                            & HF   & FCI  & FCI  & FCI  & FCI   & FCI   & VQE  & VQE  & FCI  & FCI  & FCI  & CCSD  & CCSD  & CCSDTQ \\
Hamiltonian:                & Bare & Bare & Bare & Bare & Bare  & Bare  & DUCC & DUCC & DUCC & DUCC & DUCC & Bare  & Bare  & Bare   \\
Orbitals{\footnotemark[1]}: &      & [9]  & [9*] & [15] & [15*] & [13*] & [9*] & [13*]& [9*] & [15*]& [13*]& [48*] & [48] & [48*]  \\
\hline
0  & -78.0397 & -78.0593 & -78.0593 & -78.0851 & -78.0850 & -78.0738 & -78.3063 & -78.3099 & -78.3063 & -78.3109 & -78.3099 & -78.3452 & -78.3499 & -78.3563 \\
10 & -78.0375 & -78.0569 & -78.0569 & -78.0830 & -78.0829 & -78.0716 & -78.3042 & -78.3078 & -78.3042 & -78.3088 & -78.3078 & -78.3432 & -78.3479 & -78.3543 \\
20 & -78.0307 & -78.0500 & -78.0500 & -78.0767 & -78.0766 & -78.0653 & -78.2978 & -78.3016 & -78.2978 & -78.3026 & -78.3016 & -78.3372 & -78.3419 & -78.3485 \\
30 & -78.0196 & -78.0391 & -78.0390 & -78.0662 & -78.0662 & -78.0549 & -78.2874 & -78.2913 & -78.2874 & -78.2923 & -78.2913 & -78.3272 & -78.3319 & -78.3388 \\
40 & -78.0040 & -78.0244 & -78.0244 & -78.0519 & -78.0518 & -78.0406 & -78.2729 & -78.2770 & -78.2729 & -78.2780 & -78.2770 & -78.3134 & -78.3181 & -78.3254 \\
50 & -77.9841 & -78.0063 & -78.0063 & -78.0340 & -78.0339 & -78.0227 & -78.2545 & -78.2589 & -78.2545 & -78.2598 & -78.2589 & -78.2959 & -78.3005 & -78.3085 \\
60 & -77.9599 & -77.9854 & -77.9854 & -78.0130 & -78.0129 & -78.0017 & -78.2324 & -78.2371 & -78.2324 & -78.2381 & -78.2371 & -78.2750 & -78.2796 & -78.2888 \\
70 & -77.9317 & -77.9629 & -77.9629 & -77.9902 & -77.9903 & -77.9791 & -78.2071 & -78.2123 & -78.2071 & -78.2133 & -78.2123 & -78.2512 & -78.2559 & -78.2675 \\
80 & -77.8996 & -77.9421 & -77.9421 & -77.9688 & -77.9687 & -77.9579 & -78.1794 & -78.1852 & -78.1794 & -78.1865 & -78.1852 & -78.2259 & -78.2305 & -78.2497 \\
90 & -77.8640 & -77.9353 & -77.9308 & -77.9599 & -77.9599 & -77.9493 & -78.1515 & -78.1584 & -78.1515 & -78.1600 & -78.1584 & -78.2020 & -78.2066 & -78.2387 \\
\hline\hline
\end{tabular}
\vspace{-1em}
\footnotetext[1]{
\setlength{\baselineskip}{1em} Numbers with an asterisk indicate that the atomic core orbitals were frozen in the corresponding calculation.
}
\end{center}
\label{Ethylene-Table}
\end{table*}

\begin{table*}[ht]
\caption{Energies obtained with various methods using both bare and DUCC Hamiltonians 
for the dissociation of N$_{2}$ from 0.8 to 3.5\AA.}
\begin{center}
% \resizebox{6.5in}{!}{%
\begin{tabular}{ c c c c c c c c c}
\hline\hline
             & HF   & FCI  & HF   &  FCI &  VQE & CCSD & CCSDT & CCSDTQ \\
Hamiltonian: & Bare & Bare & DUCC & DUCC & DUCC & Bare & Bare  & Bare   \\
Orbitals:    &      & [10] & [10] & [10] & [10] & [60] & [60]  & [60]   \\
\hline
0.8  & -108.4988 & -108.5177 & -108.8202 & -108.8372 & -108.8372 & -108.8556 & -108.8663 & -108.8667 \\
0.9  & -108.8399 & -108.8719 & -109.1621 & -109.1892 & -109.1891 & -109.2081 & -109.2209 & -109.2215 \\
1.0  & -108.9680 & -109.0165 & -109.2916 & -109.3305 & -109.3304 & -109.3499 & -109.3653 & -109.3664 \\
1.1  & -108.9830 & -109.0504 & -109.3089 & -109.3611 & -109.3611 & -109.3809 & -109.3996 & -109.4012 \\
1.2  & -108.9396 & -109.0287 & -109.2682 & -109.3361 & -109.3360 & -109.3561 & -109.3786 & -109.3810 \\
1.3  & -108.8684 & -108.9846 & -109.1961 & -109.2860 & -109.2858 & -109.3056 & -109.3325 & -109.3362 \\
1.4  & -108.7864 & -108.9383 & -109.1043 & -109.2272 & -109.2270 & -109.2461 & -109.2780 & -109.2834 \\
1.6  & -108.6221 & -108.8572 & -108.9106 & -109.1146 & -109.1144 & -109.1305 & -109.1757 & -109.1848 \\
1.8  & -108.4775 & -108.7952 & -108.7445 & -109.0279 & -109.0277 & -109.0390 & -109.1099 & -109.1162 \\
2.0  & -108.3575 & -108.7510 & -108.6026 & -108.9710 & -108.9708 & -108.9814 & -109.1128 & -109.0826 \\
2.2  & -108.2602 & -108.7500 & -108.4710 & -108.9369 & -108.9368 & -108.9687 & -109.1588 &  \\
2.4  & -108.1818 & -108.7465 & -108.3476 & -108.9158 & -108.9157 & -108.9927 & -109.1899 &  \\
2.6  & -108.1186 & -108.7453 & -108.2493 & -108.9047 & -108.9046 & -109.0171 & -109.2066 &  \\
2.8  & -108.0675 & -108.7446 & -108.1762 & -108.9001 & -108.9000 & -109.0329 & -109.2156 &  \\
3.0  & -108.0259 & -108.7435 & -108.1218 & -108.8987 & -108.8986 & -109.0427 & -109.2204 &  \\
3.25 & -107.9842 & -108.7438 & -108.0725 & -108.8991 & -108.8990 & -109.0501 & -109.2236 &  \\
3.5  & -107.9511 & -108.7435 & -108.0374 & -108.9003 & -108.9002 & -109.0546 & -109.2253 &  \\

\hline\hline
\end{tabular}
% }
\vspace{-1em}
% \footnotetext[1]{
% \setlength{\baselineskip}{1em} 
% }
\end{center}
\label{N2-Table}
\end{table*}

    \begin{table*}[ht]
        \caption{\label{table4}Energies obtained with the VQE method using a {bare} Hamiltonian in the {STO-3G} basis set for $N_2$ at {$r = 1$\AA}}
        \begin{center}
% \resizebox{6.5in}{!}{%

            \setlength{\tabcolsep}{4pt}

%%
%% --- Angstrom R = 1 ---
%%
            \begin{tabular}{ c   c   c      l   c   c   c }
                \hline\hline
                Ket & Initial amplitudes &  GCCSD-FCI overlap & Iterations & Final energy & GCCSD-FCI ${\Delta}$ norm \\
                \hline

% FCI energy info
%% TODO Return the former DIIS error rows
%% Formerly 0.01 & MPS w/ DMRG D16 & 1961 & 0.006410635696 & 616 \footnote{\label{unconverged-e-high}The results are unconverged (the obtained energy is too high)}  & -103.299957974434 & 1.409673270161  \\
%% Formerly 0.00 & MPS w/ DMRG D32 & 2828 & 0.000248360805 & 288\textsuperscript{\ref{partial-conv}}   & -103.322217197922 & 1.414037933858  \\
                FCI (reference)  &   &     &  &  -107.54896677   &  \\
                HF & 0         &  0.99998511 & 208  & -107.54888970 & 0.00545666  \\
                HF & MP2       &  0.99998511 & 217  & -107.54888970 & 0.00545533  \\
                HF & MPS8      &  0.99998512 & 211  & -107.54888975 & 0.00545399 \\
                HF & MPS16     &  0.99998512 & 214  & -107.54888974 & 0.00545419 \\
                HF & MPS32     &  0.99998512 & 219  & -107.54888966 & 0.00545513 \\
                MPS8  & 0      &  0.99999210 & 857  & -107.54890641 & 0.00397316 \\
                MPS8  & MP2    &  0.99999230 & 831  & -107.54890760 & 0.00392253 \\
                MPS8   & MPS8  &  0.99999282 & 831  & -107.54891098 & 0.00378811  \\
                MPS16 & 0      &  0.99999610 & 814  & -107.54893361 & 0.00278931 \\
                MPS16 & MP2    &  0.99999457 & 711  & -107.54892260 & 0.00329318 \\
                MPS16  & MPS16 &  0.99999569 & 814  & -107.54893083 & 0.00293501  \\
                MPS32 & 0      &  0.99683591 & 256  & -107.52660187 & 0.07954980 \\
                MPS32 & MP2    &  0.89380138 & 504  & -106.71534943 & 0.46086573 \\
                MPS32  & MPS32 &  0.99999311 & 1150 & -107.54891319 & 0.00371158  \\

                % FCI (reference)  & 0  &     &  &  -107.548966776712   &  \\
                % HF & 0        &  0.999985112387  & 208  & -107.548889701006 & 0.005456668044  \\
                % HF & MP2      &  0.999985119639  & 217 & -107.548889708639  & 0.005455338915  \\
                % HF & MPS8     &  0.999985126948  & 211   & -107.548889752015   & 0.005453998817 \\
                % HF & MPS16    &  0.999985125872  & 214   & -107.548889741047   & 0.005454196255 \\
                % HF & MPS32    &  0.999985120737  & 219   & -107.548889668555   & 0.005455137603 \\
                % MPS8  & 0     &  0.999992106976  & 857   & -107.548906418617   & 0.003973166036 \\
                % MPS8  & MP2   & 0.999992306870  & 831   & -107.548907600865   & 0.003922532392 \\
                % MPS8   & MPS8 &  0.999992825110 & 831   & -107.548910987473 & 0.003788110425  \\
                % MPS16 & 0     &  0.999996109860  & 814   & -107.548933613030   & 0.002789315161 \\
                % MPS16 & MP2   &  0.999994577472  & 711   & -107.548922607846   & 0.003293183135 \\
                % MPS16  & MPS16 &  0.999995692832 & 814   & -107.548930839833 & 0.002935018874  \\
                % MPS32 & 0     &  0.996835913982  & 256   & -107.526601877638   & 0.079549808519 \\
                % MPS32 & MP2   &  0.893801389412  & 504   & -106.715349430413   & 0.460865730095 \\
                % MPS32  & MPS32 & 0.999993112054 &  1150 & -107.548913192756 & 0.003711588987  \\

                \hline\hline
            \end{tabular}

%% - Explanation footer -
%% Denominator considered too small in Hessian updates - always, will not be noted
%% Partially converged (no full convergence) -> partial convergence
%% Incomplete calculation (unconverged) -> unconverged

% Rollback to the former table column length.
            \setlength{\tabcolsep}{6pt}

% }
            \vspace{-1em}
% \footnotetext[1]{
% \setlength{\baselineskip}{1em}
% }
        \end{center}
    \end{table*}

    \begin{table*}[ht]
        \caption{\label{table5}Energies obtained with the VQE method using a {bare} Hamiltonian in the {STO-3G} basis set for $N_2$ at {$r = 2.5$}\AA}
        \begin{center}
% \resizebox{6.5in}{!}{%

% Adjust the table column length.
            \setlength{\tabcolsep}{4pt}

%%
%% --- Angstrom R = 2.5 ---
%%
            \begin{tabular}{ c     c   c   l   c   c   c }
                \hline\hline

                Ket & Initial amplitudes &  GCCSD-FCI overlap & Iterations & Final energy & GCCSD-FCI ${\Delta}$ norm \\
                \hline
% -- FCI energy info --
% Reference table
                FCI (reference)    &              &                &      & -107.44040982 &     \\
                HF & 0        &  0.96597682 & 315   & -107.44020446 & 0.26085697 \\
                HF & MP2      &  0.96590547 & 271   & -107.44020084 & 0.26113031 \\
                HF & MPS8     &  0.96570857 & 351   & -107.44020678 & 0.26188327 \\
                HF & MPS16    &  0.03853423 & 313   & -105.68544505 & 1.38669806 \\
                HF & MPS32    &  0.96704249 & 266   & -107.44020092 & 0.25673918 \\
                MPS8  & 0     &  0.99979956 & 507   & -107.44035731 & 0.02002157 \\
                MPS8  & MP2   &  0.98866806 & 113   & -107.43414946 & 0.15054521 \\
                MPS8   & MPS8 &  0.99995691 & 486\footnote{\label{partial-conv}The results are partially converged (unable to achieve full convergence under calculation limits)}  & -107.44034773 & 0.00928331 \\
                MPS16 & 0     &  0.99973794 & 1191  & -107.44036718 & 0.02289364 \\
                MPS16 & MP2   &  0.99949504 & 1023  & -107.44036140 & 0.03177910 \\
                MPS16  & MPS16 & 0.99932460 & 1139  & -107.44036168 & 0.03675316 \\
                MPS32 & 0     &  0.99999898 & 31    & -107.44040735 & 0.00142449 \\
                MPS32 & MP2   &  0.99999794 & 115   & -107.44040625 & 0.00202558 \\
                MPS32  & MPS32 & 0.99999902 & 122   & -107.44040806 & 0.00139729 \\

                % FCI (reference)    &    0          &                &      & -107.440409822540 &     \\
                % HF & 0        &  0.965976820097 & 315                                      & -107.440204469327 & 0.260856971932 \\
                % HF & MP2      &  0.965905478026 & 271                                      & -107.440200844776 & 0.261130319855 \\
                % HF & MPS8     &  0.965708575615  & 351   & -107.440206786710   & 0.261883273177 \\
                % HF & MPS16    &  0.038534233500  & 313   & -105.685445054185   & 1.386698068434 \\
                % HF & MPS32    &  0.967042494896  & 266   & -107.440200922820   & 0.256739187130 \\
                % MPS8  & 0     & 0.999799568341  & 507   & -107.440357316589   & 0.020021571304 \\
                % MPS8  & MP2   &  0.988668069783  & 113   & -107.434149468644   & 0.150545210598 \\
                % MPS8   & MPS8 & 0.999956910020 & 486\footnote{\label{partial-conv}The results are partially converged (unable to achieve full convergence under calculation limits)}  & -107.440347732407 & 0.009283316251 \\
                % MPS16 & 0     &  0.999737940444  & 1191  & -107.440367189768   & 0.022893647841 \\
                % MPS16 & MP2   &  0.999495044366  & 1023  & -107.440361408440   & 0.031779101121 \\
                % MPS16  & MPS16 &  0.999324602404 & 1139 & -107.440361680118 & 0.036753165742 \\
                % MPS32 & 0     &  0.999998985402  & 31    & -107.440407359360   & 0.001424498696 \\
                % MPS32 & MP2   &  0.999997948507  & 115   & -107.440406258243   & 0.002025582787 \\
                % MPS32  & MPS32 &  0.999999023790 & 122  & -107.440408069908 & 0.001397290083 \\

                \hline\hline
            \end{tabular}

%% - Explanation footer -
%% Denominator considered too small in Hessian updates - always, will not be noted
%% Partially converged (no full convergence) -> partial convergence
%% Incomplete calculation (unconverged) -> unconverged

% Rollback to the former table column length.
            \setlength{\tabcolsep}{6pt}

% }
            \vspace{-1em}
% \footnotetext[1]{
% \setlength{\baselineskip}{1em}
% }
        \end{center}
    \end{table*}

\section{Acknowledgement}
This  work  was  supported  by  the "Embedding Quantum Computing into Many-body Frameworks for Strongly Correlated  Molecular and Materials Systems" project, 
which is funded by the U.S. Department of Energy(DOE), Office of Science, Office of Basic Energy Sciences, the Division of Chemical Sciences, Geosciences, and Biosciences.
A portion of  the calculations have been performed using the Molecular Science Computing Facility (MSCF) in the Environmental Molecular Sciences Laboratory (EMSL) at the Pacific Northwest National Laboratory (PNNL). PNNL is operated for the U.S. Department of Energy by the Battelle Memorial Institute under Contract DE-AC06-76RLO-1830. 
We also thank the Czech Ministry of Education for support of the Czech-US bilateral cooperation under the project no. LTAUSA17033.

\bibliography{ref3.bib, vqe.bib}

\end{document}